# The influence of dynamical change of optical properties on the thermomechanical response and damage threshold of noble metals under femtosecond laser irradiation

George D.Tsibidis

*Institute of Electronic Structure and Laser (IESL), Foundation for Research and Technology (FORTH), N. Plastira 100, Vassilika Vouton, 70013, Heraklion, Crete, Greece*

Email: tsibidis@iesl.forth.gr

We present a theoretical investigation of the dynamics of the dielectric constant of noble metals following heating with ultrashort pulsed laser beams and the influence of the temporal variation of the associated optical properties in the thermomechanical response of the material. The effect of the electron relaxation time on the optical properties based on the use of a critical point model is thoroughly explored for various pulse duration values (i.e. from 110fs to 8ps). The proposed theoretical framework correlates the dynamical change of optical parameters, relaxation processes and induced strains-stresses. Simulations are presented by choosing gold as a test material and we demonstrate that the consideration of the aforementioned factors leads to significant thermal effect changes compared to results when static parameters are assumed. The proposed model predicts a substantially smaller damage threshold and a large increase of the stress which firstly underlines the significant role of the temporal variation of the optical properties and secondly enhances its importance with respect to the precise determination of laser specifications in material micromachining techniques.

## I. INTRODUCTION

Material processing with ultra-short pulsed lasers has received considerable attention over the past decades due to its important technological applications, in particular in industry and medicine [1-9]. These abundant applications require a thorough knowledge of the fundamentals of laser interaction with the target material for enhanced controllability of the resulting modification of the target relief. Physical mechanisms that lead to surface modification have been extensively explored both theoretically and experimentally [10-24].

It is well-known that after irradiation with ultrashort laser pulses, the initial electron population is highly nonthermal. Experimental observations related to the nonequillibrium dynamics of electron systems in metals shows a nonthermal electron distribution which leads to delayed heating of the thermal electrons and lattice [25, 26]. Regarding modelling of the laser-matter interaction, the traditional Two Temperature Model (TTM) [27] assumes a rapidly (instantaneous) thermalisation of the electronic distribution which ignores the nonthermal character of the hot electrons. Therefore, TTM yields an overestimation of the electronic temperature which has been also confirmed by pump-probe experiments [28]. To overcome the limitations of the TTM, analysis based on the Boltzmann's transport equations [29] or revised versions of the TTM [28, 30-32] have been proposed. Nevertheless, the TTM has been used extensively to describe efficiently the thermal response and relaxation processes in various types of materials such as metals [23], semiconductors [33], and dielectrics [34] for pulse durations that are not sufficiently small.

Despite the success of the TTM, one aspect that is usually overlooked in the theoretical description of metal heating with ultrashort-pulsed laser sources is a possible temporal variation of the optical properties of the material (i.e. reflectivity and absorption coefficient). More specifically, while a dynamical change is considered for dielectrics [35] or semiconductors [10] due to the dependence of the dielectric constant on the varying carrier density, in metals, spatio-temporally invariant optical properties are used [36]. Nevertheless, it is well known that the transient variation of the dielectric constant through the temperature dependence of the electron relaxation time potentially leads to a change of the optical properties of the material during the irradiation time that needs to be evaluated [37-44]. Therefore, the, usually, constant value of the optical properties that is assumed in simulations for metals is a rather crude approximation. Although, that approximation is traditionally applied in irradiation conditions with laser beams of longer pulse durations, theoretical predictions indicate a substantial underestimation of laser energy absorption when the pulse duration is very small. More specifically, this behaviour has been previously noted where a static consideration of the optical properties led to an incorrect evaluation of the energy deposition; that was reflected on the underestimation of the thermal response of the material and the damage threshold [39, 40, 45]. Therefore, a more complete approach is necessary that will involve a rigorous consideration of a dynamic change of the optical properties during the pulse duration.

On the other hand, a lot of work has been carried out to correlate the laser beam characteristics with morphological changes [10-24]. Nevertheless, there exist still some open questions with respect to the capability of a precise estimation of the laser energy absorption and the induced effects on the thermal response of the metal that is expressed by thermomechanical effects or phase transition (melting of the material) or surface damage (mass removal). In principle, morphological surface changes at low excitation levels are strongly related to stress generation as well as whether lattice temperatures induce large stresses (i.e. that exceeds the yield stress and causes plastic deformation) [46]. The elucidation of the aforementioned issues is of paramount importance not only to understand further the underlying physical mechanisms of laser-matter interaction and electron dynamics but also to associate the resulting thermal effects with the surface response which can be used to allow a more efficient laser-based processing of the material. Therefore, there is a growing interest to reveal the physics of the underlying processes from both a fundamental and application point of view.

To proceed with the influence of electronic excitation on the morphological changes, one aspect that has yet to be explored is the correlation of the pulse width, energy deposition, optical properties changes during the pulse duration and structural changes. To this end, we present a revised version of the classical TTM (denoted as rTTM) which comprises: (i) a component that corresponds to the temporal evolution of the optical parameters during the pulse duration, (ii) a part that describes a temporally varying laser energy absorption, instantaneous electron excitation and relaxation processes, and (iii) a thermomechanical component that describes the mechanical response of the material due to material heating. For the sake of simplicity, the investigation has been focused on single shot while a similar approach could be pursued in case of repetitive irradiation (multiple pulses). Low laser fluences have been used primarily in this work to highlight the role of electron-phonon relaxation processes and emphasise on the thermomechanical response; this is due to the fact that at larger energies more complex effects such as melting or ablation occur. In that case, hydrodynamical models or atomistic simulations are required to be incorporated into a multiscale theoretical framework which could hinder the significance of the aforementioned factors while possible major morphological changes might also be attributed to other effects [10, 47, 48]. Nevertheless, prediction for the electronic and lattice temperatures at larger fluences were also performed to estimate the energy fluence at which damage occurs. It has to be noted, though, that a more rigorous investigation is required for medium or high fluences as band-filling and saturation effects [49, 50], considerable interband transitions [51, 52], two-photon interband absorption [51, 53] and impact on the relaxation times are expected to influence the predicted damage threshold values, however, such an analysis is beyond the scope of the present work.

In the following section, we present the theoretical framework used to describe the physical mechanism that characterizes laser heating of metals with ultrashort pulses. Section III explains the numerical algorithm and the adaptation of the model to gold (Au). A systematic analysis of the results and the role of the variation of the optical properties are presented in Section IV while the laser excitation and relaxation processes between electrons and lattice subsystems at small excitation levels is thoroughly investigated. The thermomechanical response is correlated with the pulse duration through the computation of the magnitude of strain fields and displacements at different values of the pulse length. A parametric study is followed in which the role of pulse duration variation in both thermal and mechanical response of the material is investigated. All theoretical results are tested against the traditional TTM (with static optical parameters) to highlight the discrepancies and analyse the magnitude of change of the optical characteristics within the laser heating time. Fluence dependence of the thermomechanical response and determination of melting/ablation thresholds are also explored and compared with experimental results. Concluding remarks follow in Section V.

## II. Theoretical model

### A. Laser Beam Profile

The laser pulse at time $t$ is described by an energy flux (in a three dimensional space characterised by the Cartesian coordinates $x,y,z$, $I(t,x,y,z)$ provided by the following expressions

$$-\frac{\partial I(t,x,y,z)}{\partial z} = \alpha(t,x,y,z) I(t,x,y,z) = W(t,x,y,z) \tag{1}$$

$$W(t,x,y,z=0) = (1-R(t,x,y,z=0))\frac{2\alpha(t,x,y,z)\sqrt{\log(2)}}{\sqrt{\pi}\tau_p} F \exp\left(-4\log(2)\left(\frac{t-3\tau_p}{\tau_p}\right)^2\right) \times \exp\left(-\frac{2(x^2+y^2)}{(R_0)^2}\right) \tag{2}$$

where $W(t,x,y,z)$ corresponds to the absorbed laser power density at time $t$, $\alpha(t,x,y,z)$ is the absorption coefficient, $\tau_p$ is the pulse duration, $F$ is the (peak) fluence and $R_0$ stands for the irradiation spot radius. It is noted that ballistic length $\lambda_{ball}$ ($\lambda_{ball}$=100nm for Au [44]) has also been taken into account. This is performed by the replacement $\alpha \rightarrow 1/(\alpha^{-1}+\lambda_{ball})$ which is incorporated to assume the impact of the ballistic movement of the hot electrons [44].

### B. Dielectric constant

There are various models that have been used to simulate the dielectric constant of the material as a function of the laser beam wavelength. Nevertheless, the following are considered as the most accurate in terms of predicting the experimentally confirmed optical parameter values at 300K:

(i) the dielectric constant of Au is modelled by means of a modified version of the extended Lorentz-Drude model proposed by Vial et al. [54]

$$\varepsilon(\omega_L) = \varepsilon_\infty - \frac{\omega_p^2}{\omega_L^2 + i\Gamma_0 \omega_L} - \frac{f\Omega_L^2}{\omega_L^2 - \Omega_L^2 + i\omega_L \Gamma_L} \quad (3)$$

where $\omega_L$ is the laser frequency (i.e. $\omega_L = 2.3562 \times 10^{15}$ rad/s for 800nm that corresponds to photon energy equal to 1.55eV [55]), while $\Omega_L$ and $\Gamma_L$ stand for the oscillator strength and the spectral width, respectively, for plasma frequency $\omega_p = 1.328 \times 10^{16}$ rad/s, $f=1.09$, $\varepsilon_\infty = 5.9673$, $\Omega_L = 4.0885 \times 10^{15}$ rad/s, and $\Gamma_L = 658.8548 \times 10^{12}$ rad/s. The damping constant $\Gamma_0$ is the reciprocal of the electron relaxation time, $\tau_e$, which is given by $\tau_e = 1/(B_L T_L + A_e(T_e)^2)$ [56], where $T_e$, $T_L$ are the electron and lattice temperatures, respectively. Values of the coefficients $A_e$ is $1.2 \times 10^7$ (s$^{-1}$K$^{-2}$) [36, 57] while $B_L$ is taken to be $0.4967 \times 10^{11}$ (s$^{-1}$K$^{-1}$), so that $\Gamma_0 = 100.0283 \times 10^{12}$ rad/s at 300K as proposed in the original model [54]. The choice of the dielectric constant expression in Eq.3 agrees well with the experimental data [58] for photon energies in the range 1.24-2.48eV (i.e. for laser wavelengths in the range 500nm-1μm, respectively) [54].

(ii) On the other hand, the computation of the dielectric constant by considering the Lorentz-Drude model with five Lorentzian terms based on the analysis of Rakic et al. (where both interband and intraband transitions are assumed) [55] describes better the experimental results for photon energies corresponding to wavelengths smaller than 500nm but there still remains some significant disagreement (see Supplementary Material)

$$\varepsilon(\omega_L) = 1 - \frac{f_0 \omega_p^2}{\omega_L^2 - i\Gamma_0 \omega_L} + \sum_{j=1}^{k=5} \frac{f_j \omega_p^2}{\omega_j^2 - \omega_L^2 + i\omega_L \Gamma_j} \quad (4)$$

where $\sqrt{f_0}\omega_p$ is the plasma frequency associated with oscillator strength $f_0$ and damping constant $\Gamma_0 = 80.521 \times 10^{12}$ rad/s at 300K [55]. The interband part of the dielectric constant (third term in Eq.4) assumes five oscillators with frequency $\omega_j$, strength $f_j$, and lifetime $1/\Gamma_j$. Values for the aforementioned parameters are given in Ref.[55] ($f_0=0.760$, $\hbar\omega_p=9.03$eV (~$1.3719 \times 10^{16}$ rad/sec), $\omega_1=0.415$eV (~$6.3050 \times 10^{14}$ rad/sec), $\omega_2=0.830$eV (~$1.2610 \times 10^{15}$ rad/sec), $\omega_3=2.9696$eV (~$4.5116 \times 10^{15}$ rad/sec), $\omega_4=4.303$eV (~$6.5374 \times 10^{15}$ rad/sec),, $\omega_5=13.82$eV (~$2.0996 \times 10^{16}$ rad/sec), $\Gamma_1=0.241$eV (~$3.6614 \times 10^{14}$ rad/sec), $\Gamma_2=0.345$eV (~$5.2415 \times 10^{14}$ rad/sec), $\Gamma_3=0.870$eV (~$1.3218 \times 10^{15}$ rad/sec), $\Gamma_4=2.494$eV (~$3.7890 \times 10^{15}$ rad/sec), $\Gamma_5=2.214$eV (~$3.3637 \times 10^{15}$ rad/sec), $f_1=0.024$, $f_2=0.010$, $f_3=0.071$, $f_4=0.601$, $f_5=4.384$).

(iii) A critical point model has been proposed together with the Drude model for the description of permitivities of Au in the 200nm-1000nm range [59]

$$\varepsilon(\omega_L) = \varepsilon_\infty - \frac{\omega_p^2}{\omega_L^2 + i\Gamma_0 \omega_L} + \sum_{j=1}^{2} A_j \Omega_j \left( \frac{e^{i\phi_j}}{-\omega_L + \Omega_j - i\Gamma_j} + \frac{e^{-i\phi_j}}{\omega_L + \Omega_j + i\Gamma_j} \right) \quad (5)$$

where $\omega_L$ is the laser frequency (i.e. $\omega_L=2.3562 \times 10^{15}$ rad/s for 800nm that corresponds to photon energy equal to 1.55eV [55]), while $\Omega_j$ and $\Gamma_j$ stand for the oscillator strength and the spectral width, respectively, for plasma frequency $\omega_p = 1.3202 \times 10^{16}$ rad/s, $f=1.09$, $\varepsilon_\infty=1.1431$, $\Omega_1=3.8711 \times 10^{15}$ rad/s, $\Omega_2=4.1684 \times 10^{15}$ rad/s and $\Gamma_1=446.42 \times 10^{12}$ rad/s, $\Gamma_2=235.55 \times 10^{12}$ rad/s. On the other hand, $\varphi_1=-1.2371$, $\varphi_2=-$ 1.0968, $A_1$=-0.26698, and $A_2$=3.0834 [59]. The damping constant $\Gamma_0$ is the reciprocal of the electron relaxation time, $\tau_e$, which is given by $\tau_e=1/(B_L T_L+A_e(T_e)^2)$ [56], where $T_e$, $T_L$ are the electron and lattice temperatures, respectively. Values of the coefficients $A_e$ is $1.2\times10^7$ (s$^{-1}$K$^{-2}$) [36, 57] while $B_L$ is taken to be $0.5372\times10^{11}$ (s$^{-1}$K$^{-1}$), so that $\Gamma_0$=108.05$\times 10^{12}$ rad/s at 300K as proposed in the original model [59].

It is evident that while the dielectric constant provided by any of Eqs.3-5 can be used to evaluate the optical properties at $\lambda_L$=800nm due to the excellent agreement with experimental results, Eq.5 will be employed in this work as it agrees well with experiment in a wider range of wavelength values (see Supplementary Material).

Despite a more accurate methodology requires the consideration of excitation-related oscillator frequencies and strength (to describe more precisely the interband transitions), the focus of the present work is on the presentation of a simple model that correlates a (temperature-dependent) variation of the optical properties with a detailed quantification of the induced thermomechanical changes. As a result, by ignoring any further enhancement of the magnitude of the dielectric constant, to a first approximation for the laser beam wavelengths studies in this work, such corrections are not assumed. The validity of the approximation will be evaluated based on the comparison with measurable quantities (i.e. damage threshold estimation). Nevertheless, a similar description of the temporal variation of the dielectric constant and the optical properties has been performed in previous studies in Ag, Cu, Au (where the $d$-bands ~2-3eV are located below the energy Fermi, well larger than the photon energy (~1.55eV) which show a good agreement with experimental observations [39, 40, 46, 60]. In addition, in the present study, the (low) fluences that are used towards highlighting the thermomechanical response of the system lead to relatively small thermal excitation energies (less than 0.3eV) (~$k_B T_e$, $k_B$: Boltzmann constant, $T_e$: electron Temperature) which suggests that the region of the electron Density of States (eDOS) affected by thermal excitations is similar to that of the free electron gas model with only $s$ electrons being excited, in principle.

In other approaches which are based on the Drude model (where the dielectric constant is described by an expression that is similar to the one used for silicon [49]), the part of the electron relaxation time due to electron-electron scattering is taken from a random-phase approximation [38, 41, 50, 52, 61]

$$\tau_{ee}^{-1} = \frac{\pi^2\sqrt{3}\omega_p}{128(E_F)^2}\frac{(\pi k_B T_e)^2+(\hbar\omega_L)^2}{\exp(-\hbar\omega_L/k_B T_e)+1} \tag{6}$$

which for $k_B T_e \gg \hbar\omega_L$, $\tau_{ee}^{-1} = \pi^4\sqrt{3}\omega_p(k_B T_e)^2/256(E_F)^2 = A_e$ $(T_e)^2$, where $E_F$=9.2eV ($E_F$: Fermi energy for Au [38]). According to Ref. [38], simulation values for the coefficient $A_e$ for Au do not provide a linear behaviour of the thermalisation time as a function of the maximum temperature due to the fact that many simplifications are performed to produce Eq.6 (i.e. screening has not been taken into account, only electrons at the Fermi edge are considered). Therefore, the coefficient $A_e$ is taken to be equal to $1.2\times10^7$ s$^{-1}$K$^{-2}$ based on the value used in previous studies [56, 47]. The computation of values for $\tau_{ee}^{-1}$ resulting from Eq.(6) and from $A_e$ $(T_e)^2$ has been performed to compare the two approaches regarding the computation of the electron-electron scattering rate (see Supplementary Material).

Similarly, regarding a more precise estimation of the optical properties in metals, it is also important to consider a detailed band structure with eDOS, especially in the medium and high-fluence regimes. Such fluences lead to laser-induced surface modifications, since the transfer of electrons from high-density $d$-bands to low-density $s(p)$-bands results in the fluence-dependent saturation of the predominant interband absorption (i.e. it also influences the intraband absorption term). This effect is characteristic for semiconductors [49], but also occurs for $d$-metals [50] and even Al at 800 nm laser wavelength [52]. The importance of saturation effects was demonstrated also in Third Harmonic Generation experiments for silver [62]. However, the predominant role of the present work aims to reveal the differences in the thermomechanical effects at *low* fluences and therefore corrections due to band-filling and induced saturation effects are not taken into account.

The dynamic character of the optical parameters (i.e. refractive index *n*, extinction coefficient *k*, absorption coefficient *α*, and reflectivity *R*) are computed through the real and imaginary part of the dielectric constant $\varepsilon_1$ and $\varepsilon_2$, respectively [63]

$$\varepsilon(\omega_L, x, y, z, t) = \varepsilon_1(x, y, z, t) + i\varepsilon_2(x, y, z, t)$$

$$n = \sqrt{\frac{\varepsilon_1(x, y, z, t) + \sqrt{(\varepsilon_1(x, y, z, t))^2 + (\varepsilon_2(x, y, z, t))^2}}{2}}$$

$$k = \sqrt{\frac{-\varepsilon_1(x, y, z, t) + \sqrt{(\varepsilon_1(x, y, z, t))^2 + (\varepsilon_2(x, y, z, t))^2}}{2}} \quad (7)$$

$$\alpha(x, y, z, t) = \frac{2\omega_L k}{c}$$

$$R(x, y, z = 0, t) = \frac{(1-n)^2 + k^2}{(1+n)^2 + k^2}$$

## C. Elasticity Equations

The mechanical response of the material is described by the differential equations of dynamic elasticity which correlate the stress and strain generation and the induced displacement as a result of the thermal expansion and the lattice temperature ($T_L$) variation [64, 65]

$$\rho_L \frac{\partial^2 V_i}{\partial t^2} = \sum_{j=1}^{3} \frac{\partial \sigma_{ji}}{\partial x^j}$$

$$\sigma_{ij} = 2\mu\varepsilon_{ij} + \lambda \sum_{k=1}^{3} \varepsilon_{kk}\delta_{ij} - \delta_{ij}(3\lambda + 2\mu)\alpha'(T_L - T_0) \quad (8)$$

$$\varepsilon_{ij} = 1/2\left(\frac{\partial V_i}{\partial x^j} + \frac{\partial V_j}{\partial x^i}\right)$$

where $V_i$ correspond to the displacements along the *x* (*i*=1), *y* (*i*=2), *z* (*i*=3) direction, while $\sigma_{ij}$ and $\varepsilon_{ij}$ stand for the stresses and strains, respectively [66]. On the other hand, the Lame's coefficients *λ* and *μ* (for Au, *λ*=111.4GPa and *μ*=27.8 GPa [67, 68]), respectively, *α'* stands for the thermal expansion of Au and $\rho_L$ is the density of the material. The Lame's coefficients *λ* and *μ* are related to the Poisson's ratio (*v*) and Young's modulus (*E*) through the relations *v= λ/(2( λ+μ))* and *E= μ (2μ+3λ)/( λ+μ)*.

## D. Generalised Energy Balance Equations

To describe the influence of the electron dynamics on the relaxation procedure and the thermomechanical response of the material, a revised version of the TTM is used that includes the transient interaction of the thermalised electron distribution with the lattice baths. Hence, the following set of equations is employed to calculate the spatio-temporal dependence of the produced thermalized electron ($T_e$) and lattice ($T_L$) temperatures of the assembly

$$C_e \frac{\partial T_e}{\partial t} = \vec{\nabla} \cdot \left( k_e \vec{\nabla} T_e \right) - G_{eL} \left( T_e - T_L \right) - \frac{\partial I(t,x,y,z)}{\partial z}$$
$$C_L \frac{\partial T_L}{\partial t} = G_{eL} \left( T_e - T_L \right) - (3\lambda + 2\mu) \alpha' T_L \sum_{j=1}^{3} \dot{\varepsilon}_{jj} \tag{9}$$

where the subscripts $e$ and $L$ are associated with electrons and lattice, respectively, $k_e$ is the thermal conductivity of the electrons [27]

$$k_e = \chi \frac{\left( \varphi_e^2 + 0.16 \right)^{5/4} \left( \varphi_e^2 + 0.44 \right) \varphi_e}{\left( \varphi_e^2 + 0.092 \right)^{1/2} \left( \varphi_e^2 + \eta \varphi_L \right)}, \quad \varphi_e = T_e/T_F, \; \varphi_L = T_L/T_F \tag{10}$$

which is valid for a wide range for low temperatures to the order of the Fermi temperature, $T_F$ ($T_F = 6.42 \times 10^4$ K for Au [69]), where $\chi = 353$ Wm$^{-1}$K$^{-1}$, and $\eta = 0.16$ [70] while $C_e$ and $C_L$ are the heat capacity of electrons and lattice, respectively, and $G_{eL}$ is the electron-phonon coupling factor. The parameters $C_e$ and $G_{eL}$ are taken from Ref. [71] by using a fitting procedure.

## III. NUMERICAL SOLUTION

Due to the inherent complexity of Eqs.(1-10), an analytical solution is not feasible and therefore, a numerical approach is pursued. Numerical simulations have been performed using the finite difference method while the discretization of time and space has been chosen to satisfy the Neumann stability criterion. Furthermore, on the boundaries, von Neumann boundary conditions are applied while heat losses at the front and back surfaces of the material are assumed to be negligible. The initial conditions are $T_e(t=0) = T_L(t=0) = 300$K, while stresses, strains, and displacements are set to zero at $t=0$. Furthermore, the vertical stress $\sigma_{zz}$ of the upper surface is taken to be zero at all times (i.e. stress free). The parameters for Au used in the simulation are summarised in Table I. The values of the laser beam parameters used in the simulation are: the (peak) fluence is equal to $F\left(\equiv \sqrt{\pi}\tau_p I_0 / \left(2\sqrt{ln2}\right)\right)$, where $I_0$ stands for the peak value of the intensity, spot radius $R_0$ (where the intensity falls to $1/e^2$) is equal to 15μm, and pulse duration values lie in the range [110fs, 8ps]. The thickness of the Au is set to 1μm. The wavelength of the beam is $\lambda_L$=800nm. We note that, the laser beam conditions are selected so that: (i) an instantaneous thermalisation of electrons ($\tau_p$>110fs) is assumed and (ii) a phase change (material melting or ablation) or plastic deformation does not occur. Hence, only elastic displacements are assumed. On the other hand, calculations for two different values of the (peak) fluence, $F$=110mJ/cm$^2$ and 140mJ/cm$^2$, have been performed in the present work.

| Parameter | Value |
|---|---|
| $C_e$ [$10^5$ Jm$^{-3}$K$^{-1}$] | Fitting [71] |
| $C_L$ [$10^6$ Jm$^{-3}$K$^{-1}$] | 2.6 [57] |
| $G_{eL}$ [$10^{17}$ Wm$^{-3}$K$^{-1}$] | Fitting [71] |
| $A_e$ [$10^7$ s$^{-1}$ K$^{-2}$] | 1.2 [36, 57] |
| $B_L$ [$10^{11}$ s$^{-1}$ K$^{-1}$] | 0.5372 (fitting) |
| $T_{melt}$ [K] | 1337 [72] |
| $T_{crit}$ [K] | 6250 [73] |
| $T_0$ [K] | 300 |
| $\alpha'$ [$10^{-6}$K$^{-1}$] | 14.2 [67, 72] |
| $E$ [GPa] | 77.97 [68] |
| $v$ | 0.4 [67] |
| $\rho_L$ [Kgr/m$^3$] | 19.3 [70, 72] |

TABLE I: Parameters for Au used in the simulations.

In principle, a common approach followed to solve problems that involve elastic displacements [66] (or hydrodynamics [10]) is the employment of a finite difference method on a staggered grid which is found to be effective towards suppressing numerical oscillations. Unlike the conventional finite difference method, time derivatives of the displacements and first-order spatial derivative terms are evaluated at locations midway between consecutive grid points while temperatures ($T_e$ and $T_L$) and normal stresses $\sigma_{ii}$ are still computed at the centre of each element. Furthermore, shear stresses $\sigma_{ij}$ are evaluated at the grid points. Numerical integration is allowed to move to the next time step provided that all variables at every element satisfy a predefined convergence tolerance of ±0.1%.

Furthermore, in the second part of this work, to explore conditions that lead to material damage, a thermal criterion is applied to determine the damage threshold. There are two scenarios that are investigated with respect to the material damage: the first one, is related to conditions that lead to a phase change (i.e. $T_L$>$T_{melt}$, where $T_{melt}$ is the temperature at which the material melts) while the second one is associated to conditions that lead to phase explosion and ablation (i.e. $T_L$>$0.9T_{crit}$, where $T_{crit}$ is the critical point temperature) [70, 74]. Both of these conditions lead to surface modification; the former induces a simple mass displacement as a result of fluid transport and solidification. By contrast, the latter scenario includes a mass removal.

We note, that in a previous work, melting and disintegration of Nickel films were explored after irradiation with ultrashort pulses by using a combined atomistic-continuum modelling [47]. Certainly, a revised model that incorporates a two temperature model and atomistic simulations is expected to allow the damage threshold estimation more precisely, however, it is outside the scope of the present study to focus on this aspect.

**IV. RESULTS AND DISCUSSION**

The theoretical model presented in the previous sections suffices to describe the thermal response of the electron and lattice subsystems. The influence of various laser beam parameters such as the (peak) fluence $F$ and the pulse duration $\tau_p$ needs to be explored, however, to correlate the laser beam characteristics with the thermal response, it is important, firstly, to evaluate whether the change of the energy of the thermalised electrons is reflected on the optical property changes of the material. As a result, it is necessary to investigate possible temporal changes of the reflectivity and absorption coefficient during the pulse duration that is expected to influence laser energy deposition and absorption. A variant energy absorption will, firstly, affect excitation and relaxation processes, and secondly, the thermal and mechanical response of the material.

Our simulations using rTTM confirm that for various pulse durations ($\tau_p$=110fs, 400fs, 2ps, and 6ps) at $F$=110mJ/cm$^2$ and at $x=y=z=0$ (where energy deposition is higher) there is a distinct variation of the magnitude of the optical properties during the pulse duration which is more pronounced at small pulse durations while at longer $\tau_p$ the discrepancy vanishes. Results are illustrated in Fig.1 and Fig.2 for reflectivity and absorption coefficient, respectively, where pronounced variations ~38% and ~54% for the reflectivity and absorption coefficient, respectively are shown for $\tau_p$=110fs while for $\tau_p$=6ps the values of the optical parameters remain almost invariant (less than 2% and 1% difference, respectively). This discrepancy increases at larger fluences (See Supplementary Material). It is emphasised that although the reflectivity varies substantially, the laser beam penetration is not expected to cause a notable effect as the ballistic length (~100nm) is much larger than the factor $1/\alpha$. Previous investigations of the transient optical properties after irradiation of noble metals (such as Au [40] or Cu [39]) demonstrate that there is a notable temporal change of both the reflectivity and the absorption coefficient that are also validated through pump-probe experiments performed by measuring reflectivity changes [44]. The noteworthy variance with decreasing $\tau_p$ is reflected on the sharp increase of the damping coefficient $\Gamma_0$ that produces remarkable changes to the electron relaxation time (Fig.3).

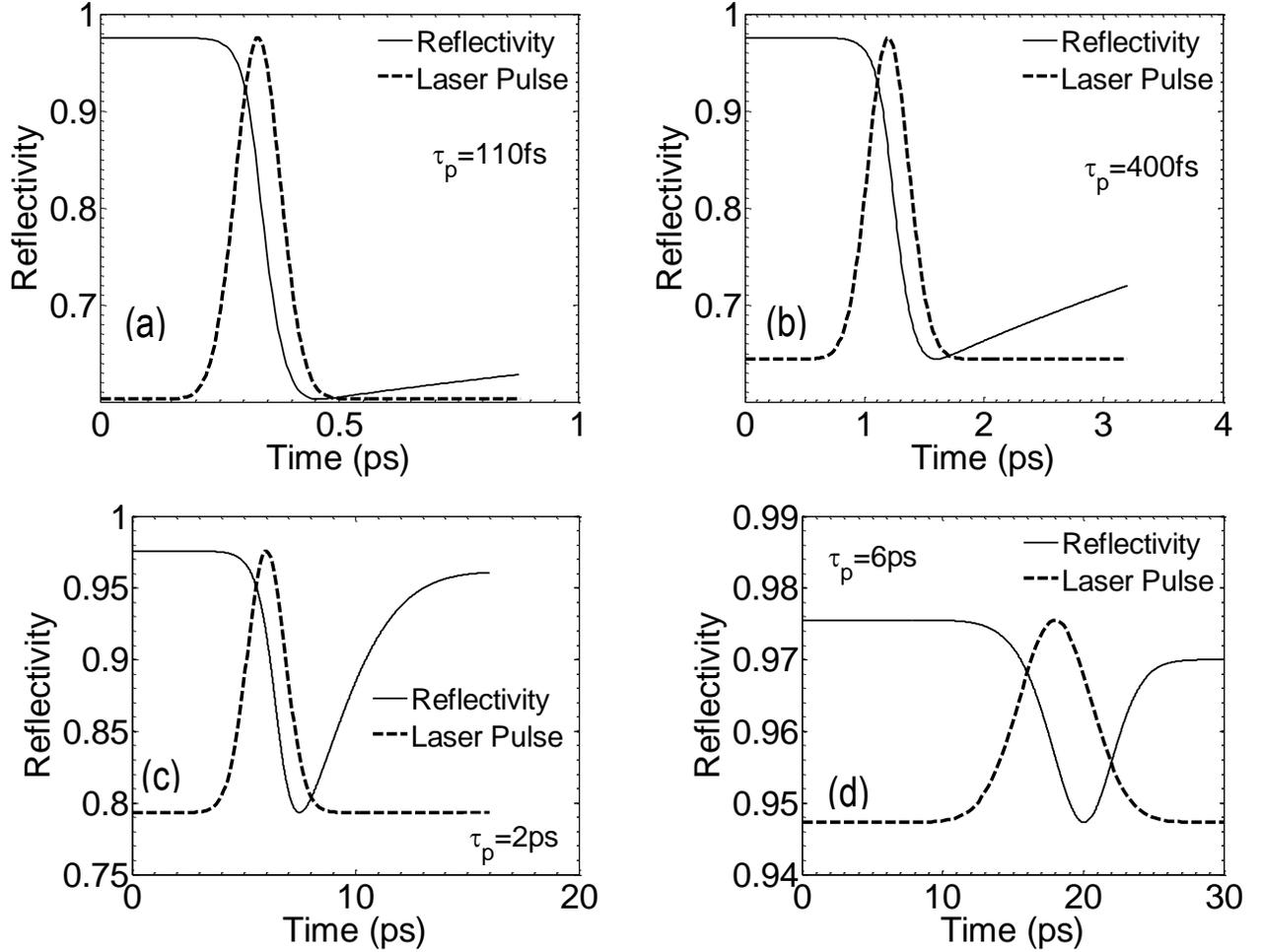

FIG. 1. Temporal dependence of the reflectivity for four different values of pulse duration ($\tau_p$=110fs, 400fs, 2ps and 6ps). ($F$=110mJ/cm$^2$, 800nm laser wavelength, $R_0$=15μm, $x=y=z=0$).

The difference between the minimum and maximum values of the optical parameters for the reflectivity and absorption coefficient as a function of the pulse duration is illustrated in Fig.4a, and Fig.4b, respectively. The predicted decrease of reflectivity at lower $\tau_p$ is expected to lead to larger energy absorption while the decrease of the absorption coefficient affects the decay length of the electromagnetic wave and the laser energy localization. To provide a detailed analysis of the optical properties variation and influence of the electron-phonon relaxation process, it is important to provide a computed estimate of the energy density per unit time absorbed by the material assuming an instantaneous excitation of the electron system. The comparison of the temporal evolution of the absorbed power densities for constant and temporally varied optical parameters (Fig.5) shows a pronounced increase of the absorbed energy at smaller $\tau_p$ if it assumed that the reflectivity and absorption coefficients are time-dependent. Therefore, the consideration of the variation of the optical properties leads to varying energy absorption in the material that is expected to induce discrepancies in the thermal response of the system and the estimation of damage thresholds [39, 40]. The solution of Eqs.1-9 allows the investigation of the thermal response of the system through the analysis of thermalisation process and the evolution of $T_e$ and $T_L$. A comparison of the maximum surface electron and lattice temperatures as a function of time (for four $\tau_p$=110fs, 400fs, 2ps, and 6.5ps) at $F$=110mJ/cm$^2$ simulated with the traditional TTM and rTTM is presented in Fig.6. Similar simulations have been performed for 140mJ/cm$^2$ (See Supplementary Material). By comparing the two models, it is evident, firstly, that the electron and lattice temperatures attained by the system (for the same $F$) are higher if a dynamical change of the optical properties is considered. This consequence is ascribed to the fact that at smaller $\tau_p$ the reflectivity decreases which leads to an increased laser energy absorption from the material and enhanced electron excitation. Therefore, the produced maximum values of $T_e$ and $T_L$ are expected to rise. On the other hand, there is a delay in the equilibration process that increases with decreasing $\tau_p$. This is due to the fact that at shorter pulses, the variation of the optical parameters is more enhanced that leads, firstly, to an increase of the absorbed energy, larger electron temperatures and a delayed relaxation process (i.e. more time is required for equilibration between the hotter electron system with the cold lattice bath). By contrast, TTM yields that the $T_e$ and $T_L$ equilibrate faster due to the smaller energies of the electron and lattice systems predicted erroneously by the invariant optical parameters. The decrease of the equilibration time as a function of $\tau_p$ predicted by rTTM and TTM is illustrated in Fig.7. It is evident that an increase of the laser fluence heats the material more efficiently which causes also an enhanced equilibration delay (Fig.7).

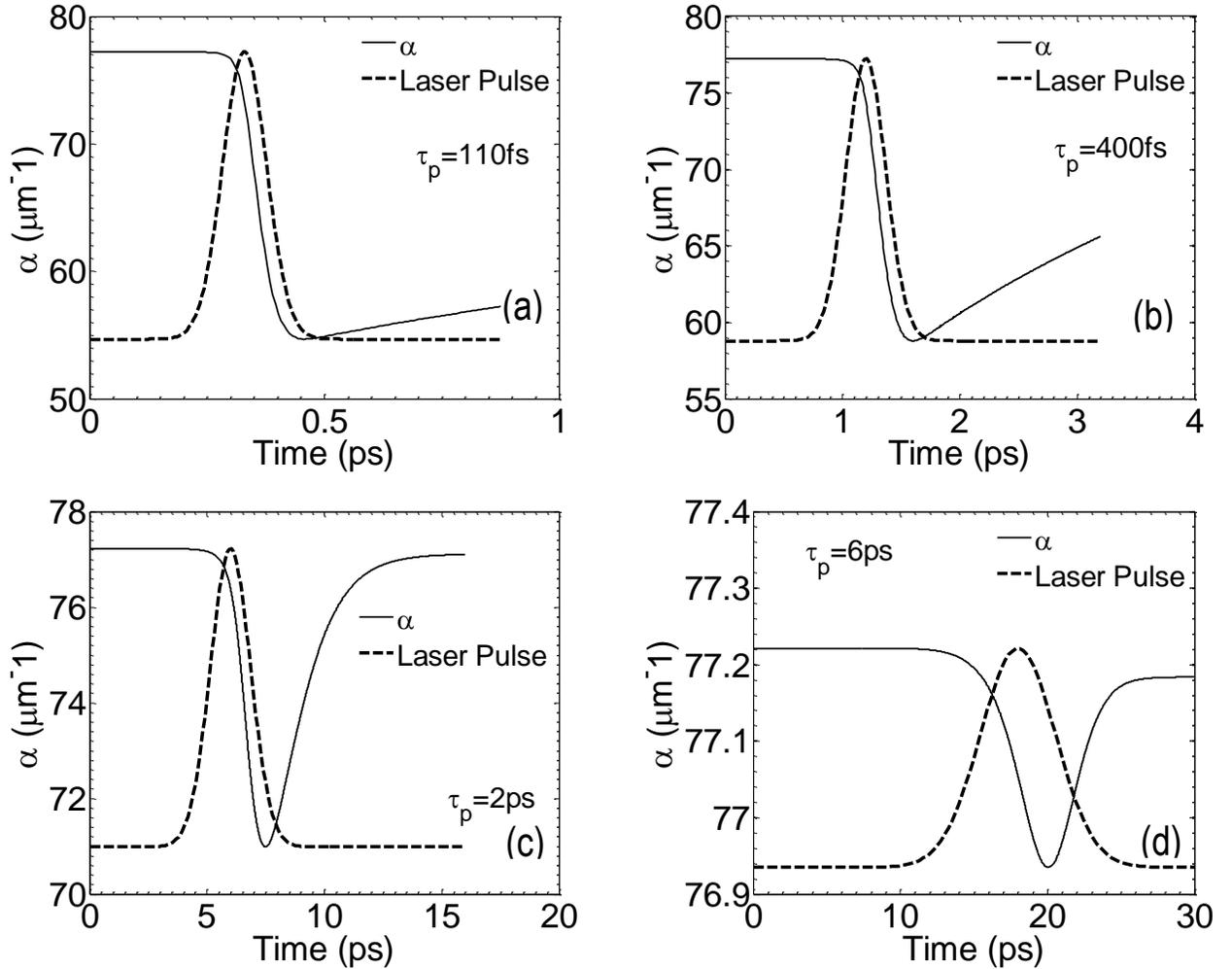

FIG. 2. Temporal dependence of the absorption coefficient for four different values of pulse duration ($\tau_p$=110fs, 400fs, 2ps and 6ps). ($F$=110mJ/cm$^2$, 800nm laser wavelength, $R_0$=15μm, $x=y=z=0$).

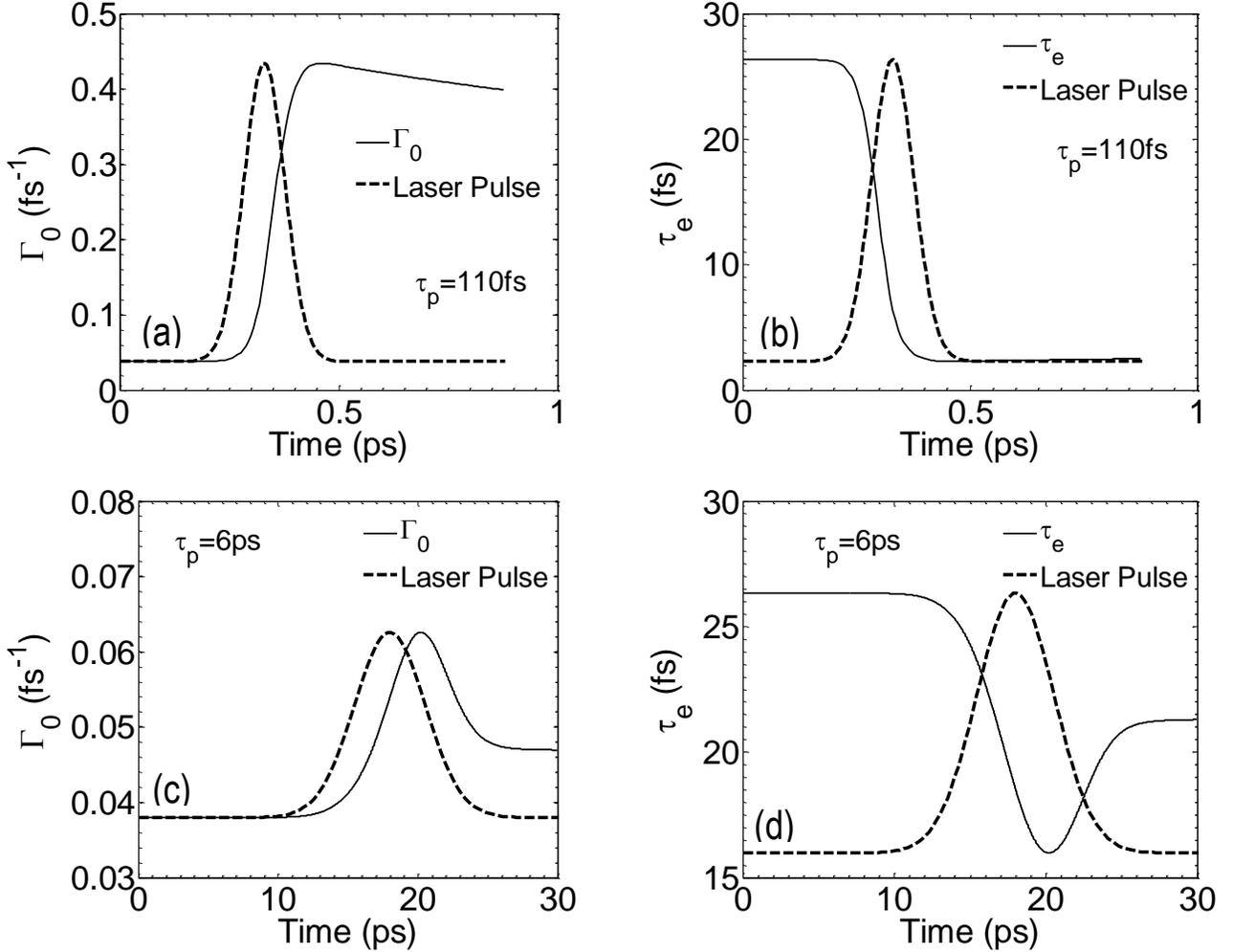

FIG. 3. Temporal variation of damping coefficient $\Gamma_0$ (a), (c) and electron relaxation time $\tau_e$ (b), (d) for $\tau_p$=110fs and 6ps, respectively. ($F$=110mJ/cm$^2$, 800nm laser wavelength, $R_0$=15μm, $x=y=z=0$).

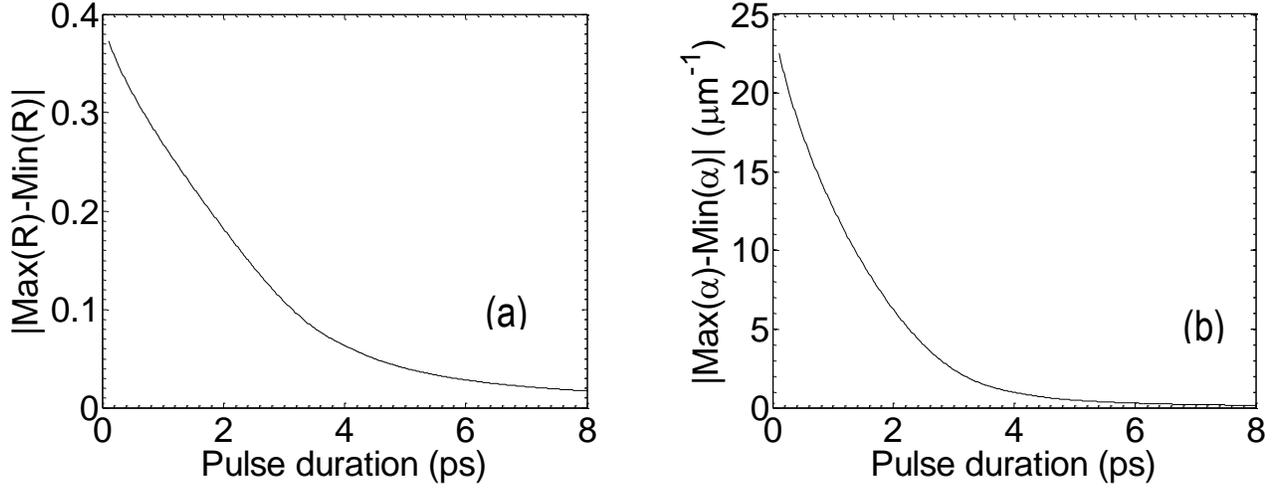

FIG. 4. Dependence of the difference of the maximum and minimum values of the reflectivity (a) and the absorption coefficient (b) on the laser pulse duration. ($F$=110mJ/cm$^2$, 800nm laser wavelength, $R_0$=15μm, $x=y=z=0$).

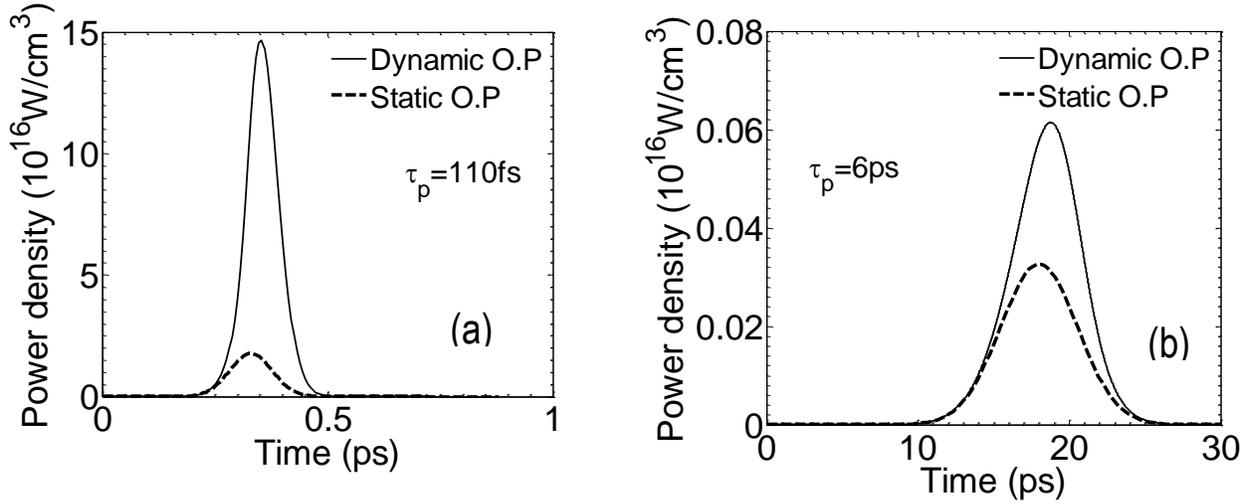

FIG. 5. Dependence of the absorbed power density temporal distributions on the pulse duration for (a) 110fs (b) 6ps, considering a static or dynamic change of the optical properties (O.P). ($F$=110mJ/cm$^2$, 800nm laser wavelength, $R_0$=15μm).

Furthermore, it appears there is also a larger discrepancy between the $T_e$ and $T_L$ values (after equilibration has been reached) if a dynamic change of the optical properties is assumed. Similar discrepancies for various fluences have been noted in previous works [24, 39]. To explain the discrepancy, one has to explore in detail the relaxation process; it is well-known that the equilibration process is dictated by two competing mechanisms: (i) electron diffusion into deeper depths inside the material and (ii) electron-phonon that governs the efficiency of localisation of energy. The pronounced role of the electron diffusion in deeper depths is reflected on the magnitude of this discrepancy (i.e. a more enhanced electron diffusion and electron heat flow $k_e \partial T_e / \partial z$ for a larger absorbed energy is predicted by rTTM (See Supplementary Material)). Thermal response of the system for various pulse duration values in the range [110fs, 8ps] for $F=110$mJ/cm$^2$ and 140mJ/cm$^2$ show a substantial decrease of the maximum electron/lattice temperature with increasing pulse duration (at the same fluence) illustrated in the graphs (Fig.8). It is also noted that the maximum lattice temperature discrepancy between the values predicted by TTM and rTTM increases for larger fluences. It is evident that the computed maximum lattice temperature is below $T_{melt}$ that shows surface damage does not occur for the laser beam conditions of the simulations. The maximum $T_L$ predicted from TTM is also almost constant which is expected as it is assumed that the same amount of laser energy is absorbed regardless of the pulse duration (Fig.8).

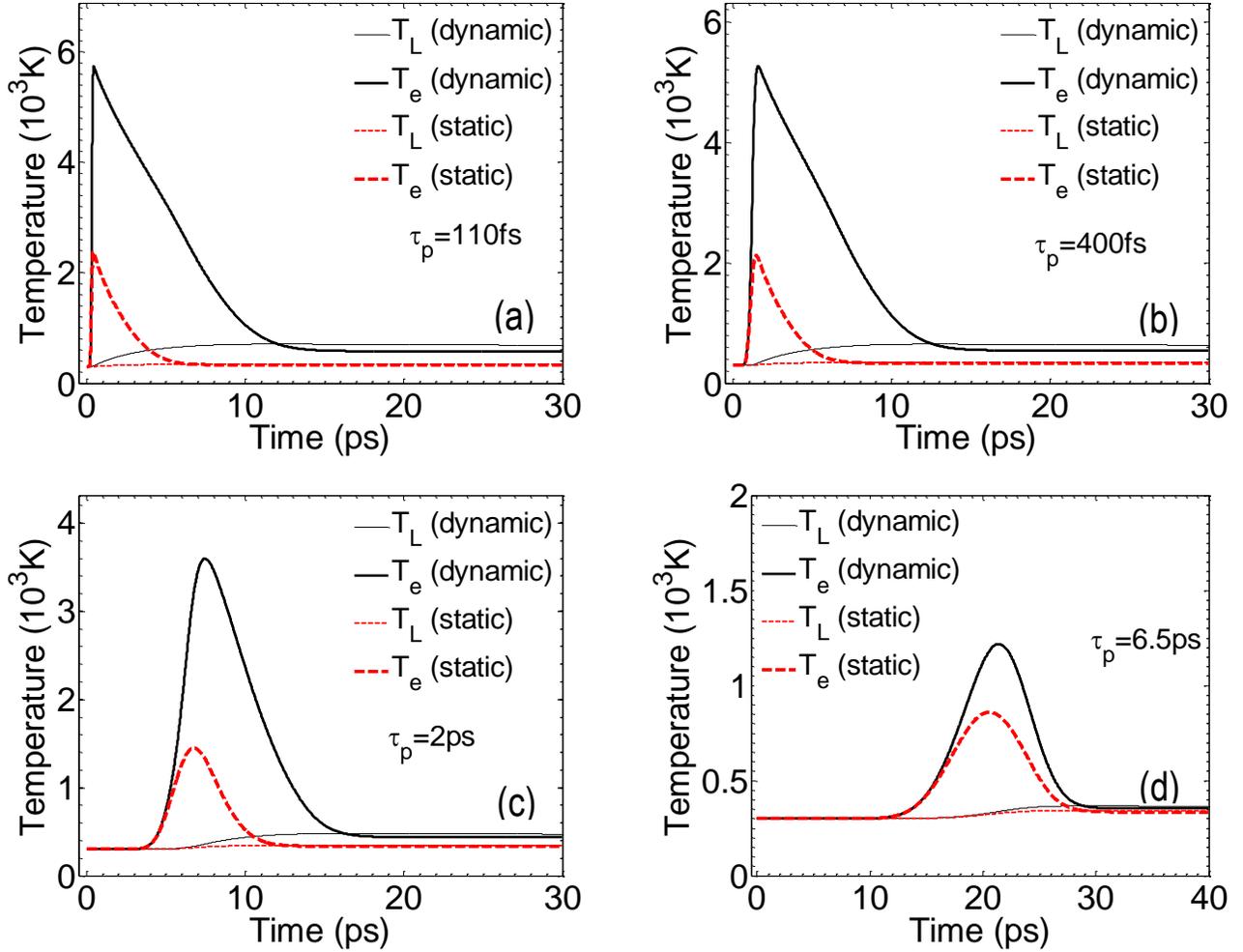

FIG. 6. Electron and lattice evolution for $\tau_p=$110fs (a), 400fs (b), 2ps (c), and 6.5ps (d) derived from rTTM and TTM. ($F=$110mJ/cm$^2$, 800nm laser wavelength, $R_0=$15μm, $x=y=z=0$).

To explore how the variation of the lattice temperature induced by the absorption of the ultrashort optical pulse leads to mechanical effects, it is important to simulate the generation of thermal stresses. It is known that thermal stresses produce strain generation and propagation. The spatio-temporal strain/stress pulse shape is determined by the solution of Eqs.7,8. To emphasise on the differences of the magnitude and spatial distribution of the strains and stresses predicted by the rTTM and TTM, the components of these fields for different $\tau_p$, along the direction of energy propagation (*z*-axis) are calculated and illustrated in Fig.9 ('Multimedia view') and Fig.10 ('Multimedia view'), respectively. The strain $\varepsilon_{zz}$ is always positive at $z=0$ due to the fact the stress-free boundary condition does not imply any stretching perpendicular to the free surface.

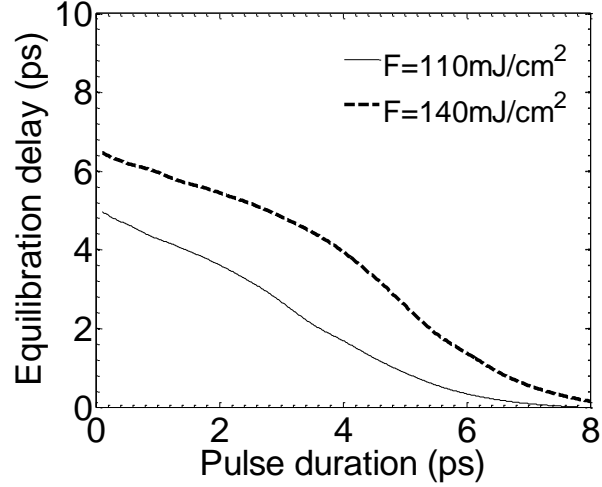

FIG. 7. Equilibration delay between predictions by rTTM and TTM for 110mJ/cm$^2$ (*solid* line) and 140mJ/cm$^2$ (*dashed* line). (at $x=y=z=0$).

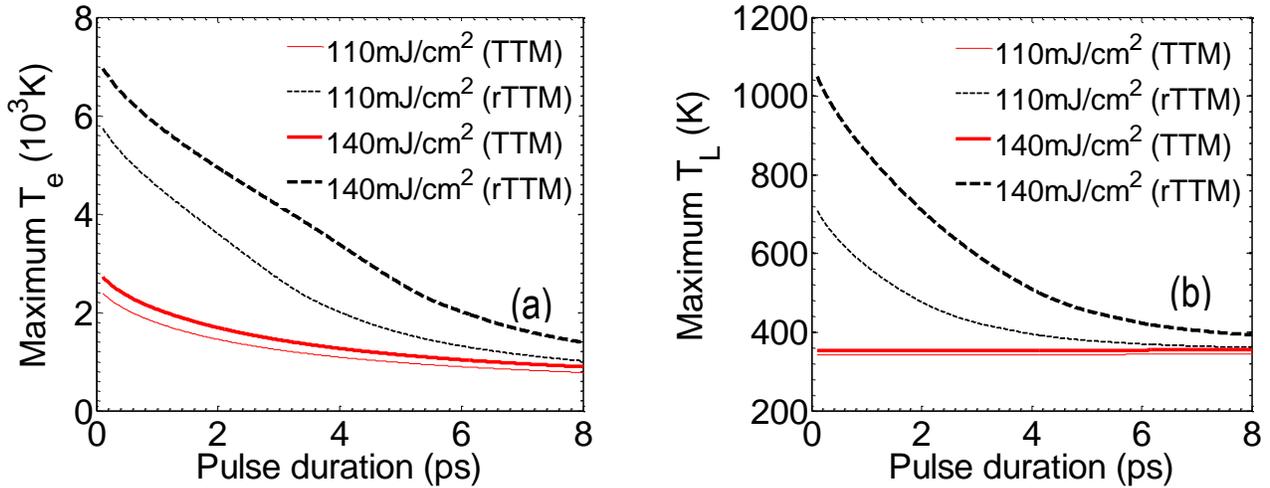

FIG. 8. Dependence of the maximum values of the $T_e$ (a) and $T_L$ (b) on the laser pulse duration assuming predictions from TM and rTTM. (800nm laser wavelength, $R_0$=15μm, $x=y=z=0$).

The solution of the first equation in Eqs.7 yields a two term propagating part, one showing a positive strain and a second with symmetric negative strain (result of the reflection on the surface). The two terms exhibit an exponential decay with length equal to $1/\alpha$. The exponential decay is reflected also on the stress fields which offer a more accurate calculation of the exponential decay length (Fig.10) ('Multimedia view'). Similar results for the strain pulse propagation have been presented in previous works for picosecond light pulses [75, 76]. Furthermore, reflectivity changes due to strain generation after ultrashort-pulsed laser irradiation of thin films on silicon surfaces have been also recently investigated [46, 60] Comparing the strain values at $z=0$ with the results predicted in previous works ([75, 76]), the non-constant $\varepsilon_{zz}(z=0,t)$ at all times is attributed to the fact that lattice temperature rise after irradiation neither occurs instantaneously nor remains constant. It is evident by estimating the spatial position of the lowest strain or stress values at different times (Figs.9,10 ('Multimedia view')) that the strain and stress pulses propagate at a speed equal to approximately, $\sqrt{(2\mu+\lambda)/\rho_L}$ =3651m/sec which is close to the experimentally measured longitudinal sound velocity in Au (~3360m/sec [77]). The pulses are illustrated at different timepoints after the arrival of the laser pulse on the surface of the material. In addition to $F$=110mJ/cm$^2$, simulations have been performed for $F$ =140mJ/cm$^2$ where similar results are deduced (See Supplementary Material).

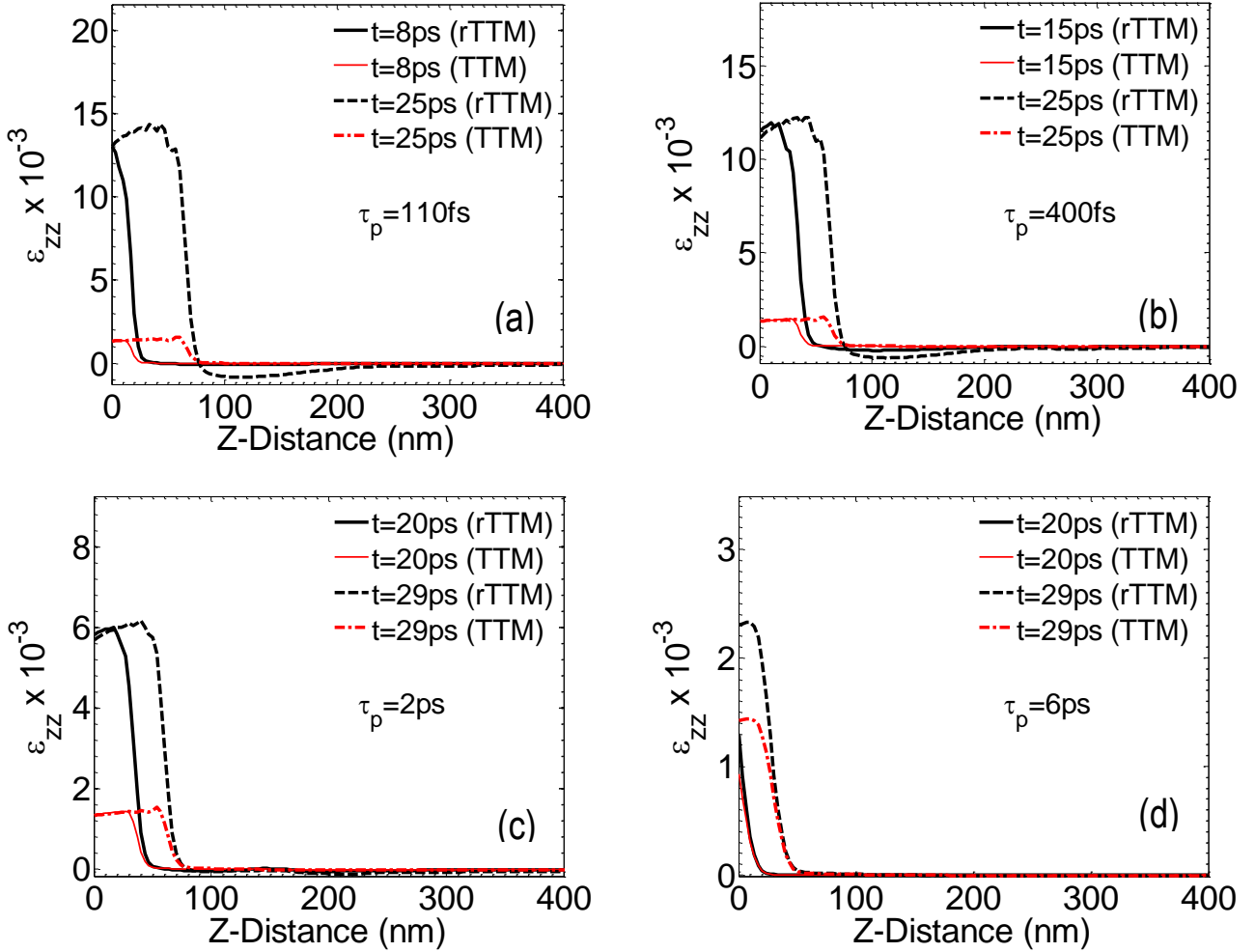

FIG. 9. Spatial dependence of strain along *z*-direction (at *x=y*=0) at various time points for $\tau_p$=110fs (a), 400fs (b), 2ps (c), and 6ps (d) derived from rTTM and TTM. ($F$=110mJ/cm$^2$, 800nm laser wavelength, $R_0$=15μm).('Multimedia view').

The comparison of the $\varepsilon_{zz}$ and $\sigma_{zz}$ pulses using rTTM and TTM demonstrates, firstly, that there is not a temporal shift of the waves as there is no delay of the lattice heating predicted by the two models (i.e. only delay in the equilibration process occurs). By contrast, the amplitude and shape of the strain and stress pulses derived from the two models appear to be significantly different. More specifically, there exists a substantial increase in both the strain and stress size in a region near the surface as a result of the significantly larger temperatures that are developed (see Fig.9 and Fig.10 and videos in Supplementary Material) ('Multimedia view'). A similarly substantially large deviation is also evident at bigger depths (i.e. it is more obvious for the stresses (Fig.10) ('Multimedia view')).

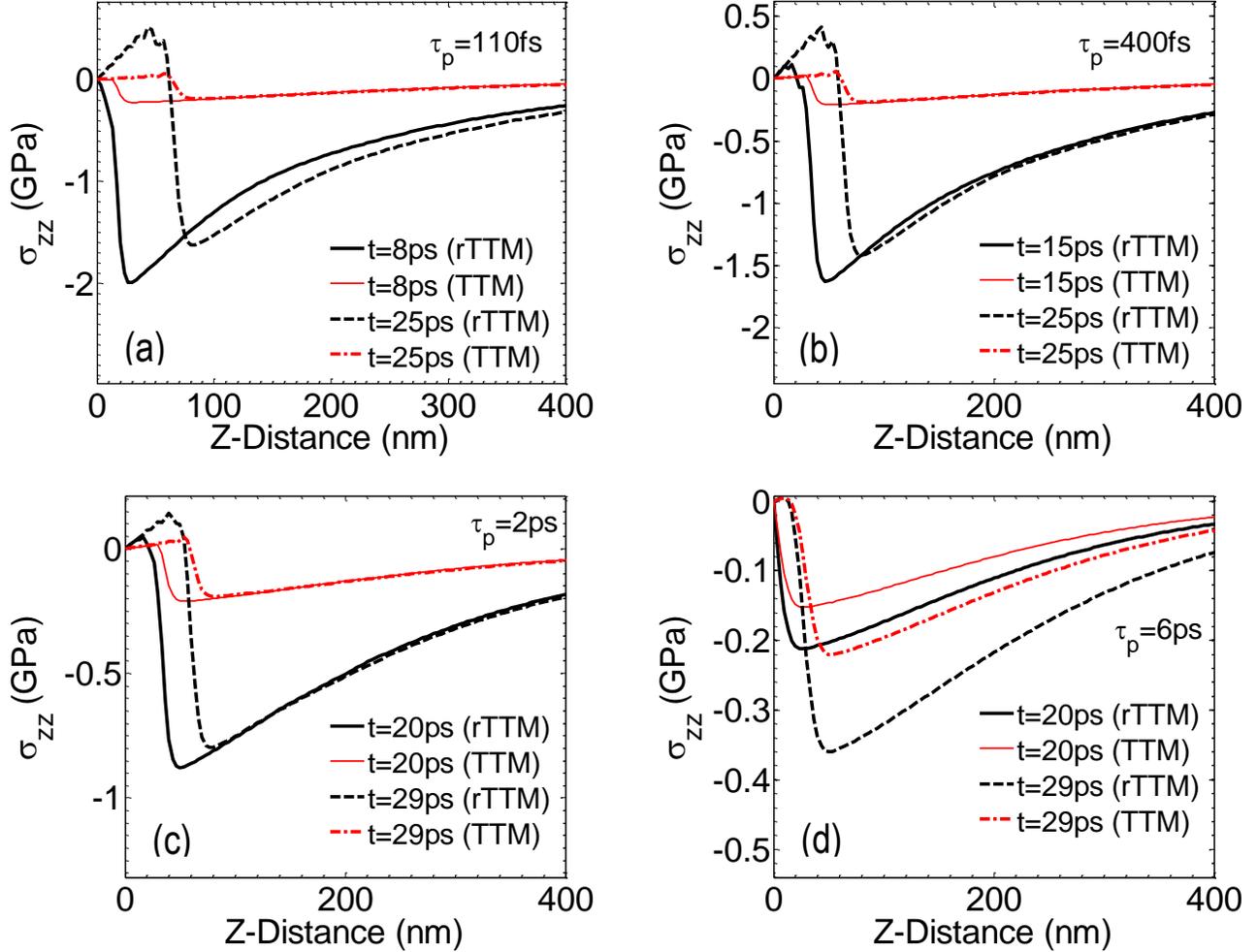

FIG. 10. Spatial dependence of stress along *z*-direction (at *x=y*=0) at at various time points for $\tau_p$=110fs (a), 400fs (b), 2ps (c), and 6ps (d) derived from rTTM and TTM. ($F$=110mJ/cm$^2$, 800nm laser wavelength, $R_0$=15μm). ('Multimedia view').

An experimental validation of the proposed mechanism is required to test the adequacy of the theoretical model with respect to the correlation of the maximum temperatures with the pulse duration and the induced thermomechanical response. However, the scope of this work is primarily related to the introduction of a consistent theoretical framework that will estimate the influence of the temporal evolution of the optical parameters to the thermomechanical response of the system due to laser heating which is expected to set the basis for an experimental confirmation. Nevertheless, a theoretical investigation of the thermomechanical response of the material and comparison of the simulation results with experimental observables in previous works where a simpler version of the model was used [46, 60] appear to confirm the adequacy of the proposed underlying physical mechanism of laser matter interaction and associated strain-generation related processes.

The aforementioned description indicates that single shot laser irradiation at low fluence influences the strain/stress fields shape and amplitude if the role of the temporal evolution of the optical properties is taken into account. Moreover, the picture is expected to become even more drastic in different laser beam conditions. For example, in multiple shot experiments of small temporal delays between the subsequent pulses (i.e. train-pulse technology), the enhanced variation of the strain wave amplitude or the substantially larger lattice temperatures produced due to excitation could influence (through accumulation effects and occurrence of plastic effects) surface micromaching techniques and applications. Furthermore, the significant deviation of the magnitude of the strain fields could also influence mechanical properties of bilayered materials (for example, thin films on silicon surfaces [46, 60]) where strong acoustic waves are expected to be reflected on the interface and interfere strongly with the propagating strain leading to a more complex total strain.

The focus of the above simulations was, firstly, to underline the differences between theoretical predictions from TTM and rTTM in conditions that do not induce material damage. Nevertheless, it is important to emphasise on the significant impact of the discrepancy between the predicted maximum lattice temperatures derived from the two models. This prospect is expected to be very significant on material properties and industrial applicability in terms of capability to modulate laser parameters as it will provide a more accurate and precise range of fluences to avoid surface damage. Therefore, in regard to the employment of the model to explore mechanisms related to onset of damage, simulations have been performed for two types of processes that account for material damage through: (i) melting and (ii) mass removal:

**(i) Melting threshold**

To estimate the fluence threshold that leads to material melting and causes damage through mass displacement, the maximum lattice temperatures predicted by both rTTM and TTM are evaluated for a range of fluences for $\lambda_L$=800nm and $\tau_p$=110fs. A comparison of the computed maximum lattice temperatures for various values of $F$ demonstrates that the maximum lattice temperature predicted using the revised model (rTTM) is always higher than the one computed by means of TTM (Fig.11). More specifically, simulation results illustrated in Fig.11 suggest that the revised model predicts a melting threshold equal to $F_{melting\ threshold}^{rTTM}$ =0.164J/cm² which is substantially smaller than the estimated by the use of the classical TTM ( $F_{melting\ threshold}^{TTM}$ ~2.81J/cm$^2$). On the other hand, a departure from the expected linear dependence of the lattice temperature on the fluence (for static $R$ and $\alpha$) is demonstrated in Fig.11 as the thermal effects are influenced by the enhanced (and non-linear) laser energy absorption. The significant difference between the above values of the melting threshold reveals the important thermal effects to the material due to the consideration of the temporal evolution of the optical properties that influences the thermal response of the system (i.e. interaction of the thermalized electrons with the lattice and lattice effects).

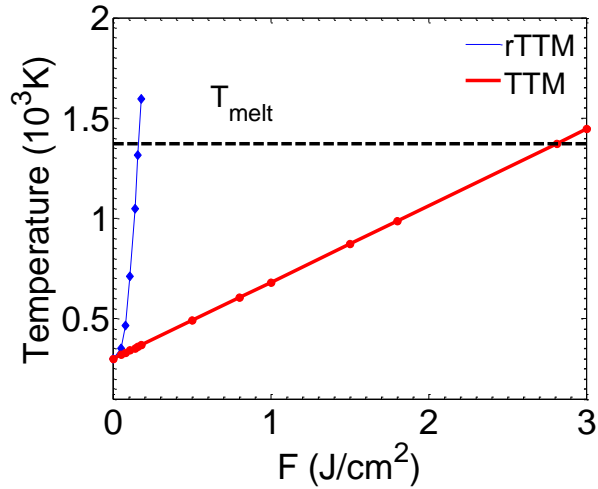

FIG.11. Maximum lattice temperatures as a function of the laser pulse fluences (800nm laser wavelength, $\tau_p$=110fs, $R_0$=15μm) simulated with TTM and rTTM. Dashed line indicates the melting temperature $T_{melt}$.

### (ii) Mass removal threshold

Simulations have also been carried out to correlate the onset of material removal with the damage threshold. The thermal criterion used to determine material removal conditions was based on the assumption that ablation onset occurs when the lattice temperature exceeds the critical temperature $0.9T_{crit}$ (=5625K) [70, 74].

### (iii) Comparison of theoretical predictions and experimental validation

To provide a validation of the theoretical framework and compare the accuracy of the two models, rTTM and TTM, the theoretical predictions for damage threshold are tested against experimental data (Table II) at different values of $\lambda_L$ and $\tau_p$. The predicted values of the ablation threshold derived from rTTM and TTM, denoted by $F_{ablation\ threshold}^{rTTM}$ and $F_{ablation\ threshold}^{TTM}$, respectively, for different Ref parameters are summarized in Table II. Similarly, calculations were performed to simulate the estimation of the melting threshold ( $F_{melting\ threshold}^{rTTM}$ or $F_{melting\ threshold}^{TTM}$ ) at different laser beam parameters (Table II).

| $\lambda_L$ [nm] | $\tau_p$ [fs] | $F_{damage\ threshold}^{exp}$ [J/cm$^2$] | $F_{ablation\ threshold}^{TTM}$ [J/cm$^2$] | $F_{ablation\ threshold}^{rTTM}$ [J/cm$^2$] | $F_{melting\ threshold}^{TTM}$ [J/cm$^2$] | $F_{melting\ threshold}^{rTTM}$ [J/cm$^2$] |
|---|---|---|---|---|---|---|
| 248 | 500 | 0.170 [78] | 0.32 | 0.29 | 0.069 | 0.067 |
| 400 | 200 | 0.025 [79] | 0.35 | 0.33 | 0.075 | 0.072 |
| 775 | 148 | 0.680 [80] | 8.73 | 0.46 | 1.78 | 0.10 |
| 785 | 130 | 0.900 [81] | 8.42 | 0.45 | 1.76 | 0.08 |
| 800 | 110 | - | 4.00 | 0.422 | 2.81 | 0.164 |
| 1053 | 600 | 0.600 [82] | 13.42 | 0.54 | 2.70 | 0.15 |

TABLE II: Calculated and measured values of the damage thresholds.

Due to the fact that there exists a controversy on the *damage* definition on the surface of the material (i.e. whether the morphological change is due to a mass removal or mass displacement), both thresholds, $F_{ablation\ threshold}$ and $F_{melting\ threshold}$ have been evaluated independently and tested with experimental observations. The comparison of the model predictions and the experimental values suggest that the revised theoretical framework yields values closer to the experimental observations if ablation is assumed to be responsible for material damage. This indicates the experimental observations assume mass removal irradiation conditions. Hence, the predicted substantial decrease of the damage threshold compared to the calculated value by TTM emphasises on the significant role of the physical processes (i.e. variation of optical properties) during irradiation which should not be overlooked.

It is noted though, that at small wavelengths (248nm and 400nm) there is not a substantial optical parameter variation which suggests that the conventional TTM is equally sufficient to describe thermal equilibration. Nevertheless, due to the expected strong interband *d*- to *s/p*- transitions for irradiation with strong photon energies (the difference of $E_d$-$E_F$ is smaller than the photon energy, $E_d$ is the *d*-band energy; the same also holds in case fluences are high enough to produce through thermal excitations a modification to the eDOS), despite the computed relatively small discrepancy of the theoretical results with the experimental observations, a more precise calculation of the electron dynamics is required to determine the evolution of the optical properties. As the interband transitions become very important, the inclusion of a number of processes, previously underestimated, have to be considered. Furthermore, it has been noted (for example, for Ti and Al) that the interband transitions lead to an interband absorption saturation and an increase of the *s*-band electrons [50, 52] that eventually influences the reflectivity. Certainly, a more accurate derivation of the dielectric constant based on first principles approach and employment of Density Functional Theory is expected to provide a more precise estimation of the values of the optical parameters [83].

Following the aforementioned analysis of results summarised in Table II, a more conclusive exploration of the role of laser beam wavelength and pulse duration in the determination of the damage thresholds (i.e. characteristic for medium and high fluences) requires a thorough investigation of the underlying physical processes. More specifically, in photoexcitation by infrared (IR) and visible range for fs/ps laser pulses, relaxation phenomena, heat conduction and (possible) two-photon interband IR-absorption are expected to influence the damage thresholds. Previous works highlighted the effect of the above processes to interpret the dependence of the damage thresholds on the pulse duration and laser beam wavelength in iron [84], copper, silver [53], aluminium and silicon [51].

Speaking about morphological changes, apart from the, obviously, simple type of surface modification, the crater formation (resulting from mass displacement or removal), other types can be derived. One very interesting type of surface modification is the formation of laser induced periodic surface structures (LIPSS) [85]. One possible candidate to explain the formation of LIPSS is the excitation of surface plasmons (SP) [10, 18]. A question that rises is whether the magnitude of the transient optical response of the irradiated material is capable to influence the periodicity of excited SP if appropriate conditions are satisfied. More specifically, in other materials such as semiconductors a sufficiently high carrier density is required to satisfy the predominant condition for SP excitation $\varepsilon_1<-1$ [10, 12, 86]. In double-pulse experiments, due to the impact of the first pulse, the carrier density can increase to a level that can further be elevated to higher values that satisfies the above condition. Then, the second pulse could interfere with the excited SP and produce a spatially (periodic) modulated energy that can lead to periodical structure formation on the surface of the material [10, 11]. Although a temporarily invariant electron density is assumed in metals, the transient reflectivity due to the, predominantly, electron temperature temporal variation is also reflected on the $\varepsilon_1$ which is also $\tau_p$ dependent (See Supplementary Material); the value of the dielectric constant is expected to influence the value of the SP wavelength (= $\lambda_L/Re\sqrt{\varepsilon/(\varepsilon+1)}$ [87], for irradiation in air), the spatial modulation of the total absorbed energy and finally the frequency of the induced periodic structures. Nevertheless, simulations predict that the variation of the SP wavelength is very small which suggests that the temporal change of the dielectric constant is not sufficient to lead to measurable changes (See Supplementary Material).

In conclusion, the adequacy of the revised model to elucidate the underlying physical processes after irradiation of Au with ultrashort-pulsed lasers suggests that the rTTM could be incorporated as a complementary module in a multi-scale framework that takes also into account electron thermalisation mechanisms as well as processes at longer time-scales (i.e. phase transitions, fluid transport, solidification, ablation, etc). Nevertheless, it needs to be emphasised that the simulations performed in this work considered pulse durations that are long enough to assume that electron thermalisation is an instantaneous process. An extension of the model to take into account the role of the presence of nonthermal electrons both in the thermalisation process and the variation of the optical properties is required based on previous approaches [28, 30, 38]. The theoretical framework can also be revised to describe more efficiently the underlying physics for medium and high fluences (where effects related to band filling, considerable interband transitions, possible two-photon interband IR-absorption and induced changes in the relaxation times should be taken into account) towards estimating more accurately the damage thresholds.

Finally, it should be stressed that although the methodology presented in this work has been applied to Au, it can also be extended to other metals.

## V. CONCLUSIONS

A detailed theoretical framework was presented that describes both the dynamical change of the optical properties of the irradiated material and the induced strains and stresses in metals with weak electron-phonon coupling constant after irradiation with ultrashort pulsed lasers. The revised TTM incorporates the relaxation processes of the thermalised electrons with the lattice while an additional component is included to describe the thermomechanical effects. A parametric analysis was performed for a range of pulse duration and fluence values that shows the differences from the predictions of the classical TTM in which both reflectivity and absorption coefficient are assumed to be static. It is demonstrated that the employment of the revised framework yields remarkably large changes for the optical parameter values as well as the induced strains which are expected to be of paramount importance for laser processing techniques. On the other hand, simulation results indicate also that the proposed underlying physical mechanism leads to a substantially lower damage threshold for the irradiated material. This a very useful aspect both a fundamental and industrial point of view towards estimating a more accurate damage threshold that is very significant for laser manufacturing approaches.

## SUPPLEMENTARY MATERIAL

See Supplementary Material that provides figures of optical properties, electron/lattice response, thermomechanical behaviour at various fluences while, at submelting conditions videos that describe the spatio-temporal distribution of $\varepsilon_{zz}$ and $\sigma_{zz}$ for $\tau_p$=110fs, 400fs, 2ps and 6.5ps are included.

## ACKNOWLEDGEMENTS

G.D.T acknowledges financial support from *LiNaBioFluid* (funded by EU's H2020 framework programme for research and innovation under Grant agreement No 665337) and *Nanoscience Foundries and Fine Analysis (NFFA)*–Europe H2020-INFRAIA-2014-2015 (under Grant agreement No 654360).